\documentclass{ws-procs9x6}

\begin{document}

\title{Chiral Dynamics from Dyson-Schwinger Equations}

\author{M. S. Bhagwat and C. D. Roberts}

\address{Physics Division, Argonne National Laboratory\\
Argonne, IL, 60439, USA}

\begin{abstract}
A strongly momentum-dependent dressed-quark mass function is basic to QCD. It is central to the appearance of a constituent-quark mass-scale and an existential prerequisite for Goldstone modes.  Dyson-Schwinger equation (DSE) studies have long emphasised these facts and are a natural way to exploit them.
\end{abstract}


\bodymatter

\vspace*{4ex}

Contemporary results from the world's experimental hadron facilities impact dramatically on our understanding of the strong interaction.\cite{arrington}  Theory requires flexible tools, which can rapidly provide an intuitive understanding of information in hand and simultaneously anticipate its likely consequences.  Models, parametrisations and truncations of QCD play this role. Prominent amongst these are the DSEs and truncations thereof, and herein we will intimate recent progress.

The gap equation is primary.  Its dynamical chiral symmetry breaking solution (DCSB) only has a chiral expansion on a measurable domain of current-quark mass.\cite{Chang:2006bm}  Within this domain a perturbative expansion in the current-quark mass around the chiral limit ($\hat m=0$) is uniformly valid for physical quantities.  That fails at $\hat m_{\rm cr} \sim 60$--$70\,$MeV; i.e., a mass $m_{0^{-}} \sim 0.45\,$GeV for a flavour-nonsinglet $0^{-}$ meson constituted of equal mass quarks.

For chiral dynamics it is singularly important that at a nonperturbative, systematic and symmetry preserving DSE truncation exists.\cite{bhagwatvertex}  That enables proof of Goldstone's theorem in QCD\cite{mrt98} and yields 
\begin{equation}
\label{gengmor}
f_{\pi_n} \, m_{\pi_n}^2 = 
2 m(\zeta^2)\,\rho_{\pi_n}(\zeta^2) \,,
\end{equation}
wherein $f_{\pi_n}$ is the leptonic (pseudovector) decay constant for a pseudoscalar meson and $\rho_{\pi_n}$ is the pseudoscalar analogue.  ($n=0$ denotes the lowest-mass pseudo\-sca\-lar and increasing $n$ labels bound-states of increasing mass.)  The Gell-Mann--Oakes-Renner relation is a corollary of (\ref{gengmor}).  Importantly, (\ref{gengmor}) is equally valid for a meson containing one heavy-quark; i.e., a heavy-light system,\cite{Ivanov:1998ms} and also heavy-heavy mesons.\cite{mindthegap}  Equation\;(\ref{gengmor}) entails \cite{Holl:2004fr} that for $\hat m = 0$, 
$f_{\pi_n}^0 \equiv 0\,, \forall \, n\geq 1\,$;
viz., Goldstone modes are the only mesons to possess a nonzero leptonic decay constant in the chiral limit when chiral symmetry is dynamically broken.  The decay constants vanish for all other pseudoscalar mesons on this trajectory, e.g., radial excitations.  NB. In the absence of DCSB, the leptonic decay constant of \emph{all} such pseudoscalar meson vanishes when $\hat m = 0$; namely, $f_{\pi_n}^0 \equiv 0\,, \forall \, n\geq 0\,$.  The consequences of the axial-vector Ward-Takahashi identity for two-photon decays of pseudoscalar mesons have been explored\cite{Holl:2005vu}.  Amongst other things it has been proven that when $\hat m = 0$ the leading-order ultraviolet power-law behaviour of the transition form factor for excited state pseudoscalar mesons is O$(1/Q^4)$. 

The study of scalar mesons is naturally important.  Modern DSE analyses of the $u$,$\,d$-quark scalar meson provide results that are compatible with a picture of the lightest $0^{++}$ as a bound state of a dressed-quark and -antiquark supplemented by a material pion cloud.\cite{Holl:2005st}  Meson, baryon and dressed-quark $\sigma$-terms have also been studied.\cite{Holl:2005st,Flambaum:2005kc}  Consequences of symmetries are manifest and the behaviour of the dressed-quark $\sigma$-term shows that the essentially dynamical component of chiral symmetry breaking decreases with increasing current-quark mass.

\noindent{\bf Acknowledgments.} This work was supported by US Department of Energy, Office of Nuclear Physics, contract no.\ DE-AC02-06CH11357.

\end{document}